\documentstyle[12pt]{article}
\newcommand{\be}[1]{
\begin{eqnarray}\label{#1}}
\newcommand{\ee}{\end{eqnarray}}

  \def\beq{\begin{equation}}
  \def\eeq{\end{equation}}
  \def\beqr{\begin{eqnarray}}
  \def\eeqr{\end{eqnarray}}
  
\begin{document}
\renewcommand{\thefootnote}{\fnsymbol{footnote}}
\begin{flushright}
\begin{tabular}{l}
\\
TPR-01-05\\
\end{tabular}
\end{flushright}
\begin{center}
{\bf\Large Crossing and Radon Tomography for Generalized 
Parton Distributions}

\vspace{0.5cm}
O.V. Teryaev$^{a,b}$
\begin{center}
{\em
$^a$ Institut f\"ur Theoretische Physik, Universit\"at Regensburg \\
D-93040 Regensburg, Germany \\
$^b$ Bogoliubov Laboratory of Theoretical Physics, JINR,
141980, Dubna, Russia}
\end{center}
 \end{center}
 \vspace{1.5cm}

 \begin{abstract}

The crossing properties of the matrix elements of non-local operators,
parameterized by Generalized Parton Distribution, are considered.
They are  especially simple 
in terms of the Double Distributions which are common for the 
various kinematical regions. 
As a result, 
Double Distributions may be in principle extracted from the 
combined data in these  
regions by making use of the inverse Radon transform,
known as a standard method in tomography. The ambiguities analogous 
to the ones for the vetor potential in the  
two-dimensional magneto-statics are outlined.
The possible generalizations 
are discussed. 

\end{abstract}

\section*{\normalsize \bf Introduction}

Deeply Virtual Compton Scattering (DVCS) \cite{DVCS1,DVCS2} is the
cleanest hard process which is
sensitive to the
Generalized Parton Distributions (GPD), the most popular version of which is
probably represented by the Skewed Parton Distributions (SPD), and has been
the subject of extensive theoretical investigations for a few years.
First experimental data became recently available (see e.g. \cite{exp1,exp2})
and much more data are expected from JLAB, DESY, and CERN in the near future.
In the present note we shall limit ourself to the case of the 
pion target, although our consideration 
can be applicable to any target.  

The crossing version is then provided
by the process $\gamma^\ast \gamma \to \pi \bar \pi $ with a highly virtual
photon but small hadronic invariant mass $ W$.
It was recently investigated in the
framework
of QCD factorization\cite{DGPT,DGP,fac}. It allows to study the pion pair
produced in the isoscalar channel,  
where the huge $\rho$-meson peak is absent. This
process is analogous to the single pion production, described by the
pion transition form-factor, being the long time object of QCD studies
\cite{BL,RR}.  In particular, the generalized
distribution amplitude (GDA), describing the non-perturbative stage
of this process, is a natural counterpart of the pion light cone
distribution amplitude. From the other side, it may be considered as   
a crossing counterpart of pion SPD \cite{DGPT} and related to it
\cite{PW} by making use of the suitable polynomial basis. 

The alternative (and quite elegant) description of non-perturbative 
stage of DVCS and other hard exclusive processes is provided 
by Double Distributions (DD) \cite{DVCS1,Rad97}. They naturally explain the 
polynomiality of SPD and can be a good starting point for the model-building. 
At the same time, it is not clear how one can express 
DD in terms of the hadronic matrix element and extract them, at least 
in principle, from the data. 

In this note we study the crossing properties of the double 
distributions. We conclude, that both mentioned 
channels may be described by the 
common DD, while SPD and GDA may be obtained by its integration 
over the straight lines  constituting the various angles with the
coordinate axes. This would allow us to recover the DD from 
the SPD and GDA by making use of the inverse Radon transform -
the mathematical tool widely used for computer tomography.
We discuss the ambiguities resulting from the presence of the so-called 
Polyakov-Weiss (PW) terms in DD and 
find their close analogy with that for the recovering of the vector 
potential from the magnetic field.  
We also outline the possible applications 
of the method and its further development.

\section*{\normalsize \bf Double distributions in the crossing 
related channels}

Let us start with the following symmetric representation of the matrix 
elements of twist-2 non-local operators (c.f. \cite{PW})

\be{Parm}
\langle p'| \bar \psi\biggl(-\frac{z}{2}
\biggr) \gamma \cdot z 
\psi\biggl(\frac{z}{2} \biggr)|p\rangle = 
(2P \cdot z)
\int_{-1}^1 dx 
\int_{|x|-1}^{1-|x|} dy e^{-ix Pz - i y \Delta z/2 }
F (x,y,\Delta^2) \\ \nonumber
+(\Delta \cdot z)
\int_{-1}^1 dx 
\int_{|x|-1}^{1-|x|} dy e^{-ix Pz - i y \Delta z/2} G (x,y,\Delta^2); \\ 
\nonumber
\langle p',-p| \bar \psi\biggl(-\frac{z}{2}
\biggr) \gamma\cdot z 
\psi\biggl(\frac{z}{2}n\biggr)| 0 \rangle = 
(2P \cdot z)
\int_{-1}^1 dx 
\int_{|x|-1}^{1-|x|} dy e^{-ix Pz - i y \Delta z/2 }
f (x,y,\Delta^2) \\ \nonumber
+(\Delta \cdot z)
\int_{-1}^1 dx 
\int_{|x|-1}^{1-|x|} dy e^{-ix Pz - i y \Delta z/2} g (x,y,\Delta^2);
\ee
where we (in order to express the crossing properties in the most simple way)
adopted the common notations for both channels: 
$P=(p+p')/2,
\Delta= p'-p = 2 \xi P +\Delta_\perp$, while $p,p'$ are the
initial and final momenta of pion in the DVCS channel, respectively.
In the  $\gamma^\ast \gamma \to \pi \bar \pi$ channel, 
$\Delta$ corresponds to the total momentum of pion pair while 
$2P$ is the relative momentum of pions. 

The skewedness parameter 
varies in the region $0 < \xi < 1$ for the DVCS channel and 
$|\xi|>1$ for the  $\gamma^\ast \gamma \to \pi \bar \pi$ channel, 
respectively. In what follows we restrict ourselves to the leading 
twist level, neglecting the transverse momentum $\Delta_\perp$.

The crossing symmetry states that in the different 
channels the matrix elements are described by the common analytical functions,
so that one may relate them by analytical continuation. 
One should note the crucial advantage of the DD for the studies of 
crossing properties, namely, that $\xi$ does not enter explicitly 
to their definition.
The only variable,
which should be continued for DD is therefore $\Delta^2$ which 
is equal to t in the DVCS channel and to the squared  invariant mass 
of the pion pair $W^2$ in the $\gamma^\ast \gamma \to \pi \bar \pi $ channel. 
The dependence on these variables is entirely non-perturbative and is a 
subject of separate investigation. 
This dependence is beyond the scope of our studies,
and we shall put   
both $t$ and $W^2$ to zero. 
This point 
is the common unphysical point for the both channels, 
so one may conclude, that

\be{cross}
F (x,y,0))=f(x,y,0); \ G(x,y,0)=g(x,y,0)
\ee

The Skewed Parton Distribution are related to the Double ones by the 
integration over straight lines \cite{Rad97}. 
The symmetric skewed distribution is related to symmetric double 
distributions F and G by the formula, being the straightforward 
generalization of \cite{Rad97}.

\be{ParmH}
H(z,\xi)=\int_{-1}^1 dx 
\int_{|x|-1}^{1-|x|} dy (F(x,y) + \xi G(x,y)) \delta (z-x-\xi y).
\ee

In turn, the Generalized Distribution Amplitudes, describing the 
non-perturbative stage of hadron pair production, may be related 
to the respective Double Distributions in an analogous way: 

\be{ParmP}
\Phi (t,1/\xi)=\int_{-1}^1 dx 
\int_{|x|-1}^{1-|x|} dy (g(x,y)+\frac{f(x,y)}{\xi}) \delta (t-x/\xi - y)
\ee

Note the difference with the original definition of GDA \cite{DGPT},
where skewedness was defined by the fraction of the pair momentum carried by
the  one of the produced pions, while here it is defined (in the more
symmetric way) by the difference of their momenta.  
As a result, the analytic continuation between $\Phi$ and $H$ is provided
by the formula:

\be{contH}
H(z,\xi)=  sign(\xi) \Phi (\frac{z}{\xi}, \frac{1}{\xi})
\ee  

Contrary to (\ref{cross}), SPD and GDA, which  are depending on $\xi$
explicitly, correspond to the different regions of this variable: 
$\xi < 1$ for SPD and  $|\xi| > 1$ for GDA. In what follows, 
we shall consider SPD in the 'extended' region, on the whole real 
axis, assuming that for  $|\xi| > 1$ it should be substituted 
by GDA from (\ref{contH}), while  the region 
 $-1<\xi < 0$ is accessible
by the use of symmetries. 

These discrete symmetries of the SPD and GDA may be now described 
in the unified way. Namely, the symmetries of DD with respect
to its second argument  
\be{symD}
F(x,y)=F(x,-y) \ ; \ G(x,y)=-G(x,-y),
\ee 
correspond to either T-invariance for SPD, or charge conjugation 
invariance for GDA. 
This is not surprising, as the T-invariance, first studied 
in the case of the forward twist-3 matrix elements \cite{ET84}, 
corresponds to the interchange of both hadrons and partons, 
in complete analogy with charge conjugation.   
Note that the reality of the scalar function (SPD or DD)
plays the crucial role in such a derivation. At the same time, 
GDA may acquire the non-trivial imaginary phases due to 
cut in $W^2$. However, it is only the real part of GDA which 
is described by crossing (\ref{cross}), while the imaginary parts 
requires the separate study of the analytical continuation in  $W^2$.
Also, as soon as the chosen definition of GPD preserves P -invariance,
such a similarity of T- and C- invariance is related to CPT-theorem. 
Note that the action of the T transforms GDA 
to another objects: $ \langle 0| \bar \psi\biggl(-\frac{z}{2}
\biggr) \gamma\cdot z 
\psi\biggl(\frac{z}{2}n\biggr)|-p',p\rangle$, describing the 
appearance of $h \bar h$ pair in the initial, rather than in the final 
state. Unfortunately, they, strictly speaking, 
cannot be accessed in the process 
of  $h \bar h$-annihilation to the  
$\gamma^\ast \gamma$ pair, being the crossing version 
of DVCS and  $h \bar h$ production, as the mass of virtual photon,
which should provide the 'hard' scale, 
is constrained by the kinematics to be smaller than $s=W^2$,
which is assumed to be the 'soft' parameter. 
At the same time, one may consider the  $h \bar h$   
(in reality, probably $p \bar p$)
annihilation to the pair of two {\it real} photons at 
(moderately) large $s$ as a crossing version of the large angle
(real) Compton scattering, which is known to be described by GPD \cite{Radt}. 
This may give access to the $p \bar p$ GDA, very difficult otherwise.


The relations
(\ref{symD})
lead to the following 
symmetry properties for SPD and GDA

\be{symH}
H(x,\xi)=H(x,-\xi); \Phi(z, \xi)=-\Phi(-z,-\xi),
\ee 

The additional symmetry properties emerge if the pions are in the 
definite isospin state (where we assume that more general symmetry 
(\ref{symD},\ref{symH}) also holds)

\be{symI}
F^{I=0}(x,y)=-F^{I=0}(-x,y); G^{I=0}(x,y)= G^{I=0}(-x,y), \nonumber \\
F^{I=1}(x,y)=F^{I=1}(-x,y); G^{I=1}(x,y)= -G^{I=1} (-x,y) \nonumber \\
H^{I=0}(x,\xi)=-H^{I=0}(-x,\xi); \Phi^{I=0}(z, \xi)=-\Phi^{I=0}(-z,\xi)
=-\Phi^{I=0}(z,-\xi) \nonumber \\
H^{I=1}(x,\xi)=H^{I=1}(-x,\xi); \Phi^{I=1}(z,\xi)=\Phi^{I=1}(-z,\xi)
=-\Phi^{I=1}(z,-\xi).
\ee 
For simplicity, we shall limit ourselves for the time being to the case of 
DD $F(x,y)$ only, neglecting the Polyakov-Weiss  
\cite{PW} term, resulting from the function $G(x,y)$. This approximation
is self-consistent in the case of $I=1$ \cite{PW}.

\section*{\normalsize \bf Double distributions from the 
inverse Radon transform}

We now make the key observation, that  the relation between SPD and 
DD is nothing else than a particular case of the Radon transform
\cite{Radon1,Radon2,Radon3,Radon4}. 
As soon as all the possible lines crossing the compact 
region are considered, one is dealing with the mapping between functions of 
two variables (because each line is characterized just by two real numbers).  
As soon as the transformed function is continuous, it may be recovered 
by making use of the inverse Radon transform. This procedure is 
known to be the key ingredient of the numerous applications \cite{Radon3}, 
covering, say, medicine, optics and geophysics. We are
going to suggest its application in (non-perturbative) QCD. 

As soon as Radon transform is a new instrument in this field, it makes 
sense to present briefly the derivation of its inversion, which is actually 
very simple. It is more convenient to present it by parameterizing the 
straight line in a slightly different way: by using the unit vector 
$\vec \xi=(cos \phi, sin \phi)$ orthogonal to it, 
instead of skewedness $\xi$, and its distance from the 
origin $p$ instead of the argument of the skewed distribution $z$. 
The correspondence to (\ref{ParmH}) is obvious. 
   
\be{ParmR}
R (p,\vec \xi)=\int_{-1}^1 dx 
\int_{|x|-1}^{1-|x|} dy f (x,y) \delta (p - \vec x \vec \xi), \\
\xi=tg \phi, z=p/cos \phi; H(z,\xi)=R (p, \vec \xi) |cos \phi|.
\ee 
The most convenient way to invert the transform is to use its relation to 
Fourier transform. Indeed, the Fourier transform of the function 
$f(\vec x) \equiv  f(x,y)$ may be written in the form 

\be{Four}
F(\vec q)=
\int d^2 \vec x e^{i \vec x \vec q}
f (\vec x) = 
\int_{-\infty}^{\infty} dt \delta(t - \vec x \vec q)
\int   d^2 \vec x  e^{i \vec x \vec q}
f (\vec x)
\ee
Here the integration is performed in the whole $(x,y)$ plane,
while the correct integration limits are provided 
by the zero value of the function $f$ outside region (\ref{Parm}).
It is instructive to specify the direction of $\vec q$ by the 
unite vector $\vec \xi$, which would be shown in a moment 
to be an argument of the Radon transform

\be{FR}
F(\vec \xi \lambda)=
\int_{-\infty}^{\infty} dt \delta(t - \lambda 
\vec x \vec \xi)
\int d^2 \vec x 
e^{i \lambda \vec x \vec \xi}
f (\vec x) =_{t \to \lambda t} 
\int_{-\infty}^{\infty} dt \delta(t - \vec x \vec \xi)
\int d^2 \vec x e^{i \lambda \vec x \vec \xi}
f (\vec x)= \nonumber \\
\int_{-\infty}^{\infty} dt e^{i \lambda t}
 \int d^2 \vec x \delta(t - \vec x \vec \xi) f (\vec x)=
\int_{-\infty}^{\infty} dt e^{i \lambda t} R(t,\vec \xi)
\ee
As a result, the two-dimensional Fourier transform may be represented as
a combination of one dimensional Fourier transform and Radon transform
\footnote{This is actually the core of the relation between 
DD and SPD, being, respectively, the two-dimensional and one-dimensional 
Fourier transforms of the same matrix element.}. 
Consequently, inverting the two-dimensional Fourier transform  in 
the polar coordinates
and making use of (\ref{FR}), 
the function may be expressed through its Radon transform  

\be{FourR}
f(\vec x)=
\frac{1}{(2 \pi)^2}
\int d^2 \vec q e^{-i \vec x \vec q}
F (\vec q) = 
\frac{1}{(2 \pi)^2}
\int_0^\infty \lambda d \lambda \int_0^{2\pi}
d \phi e^{-i \lambda \vec x \vec \xi}
F (\lambda \vec \xi) = \nonumber \\
\frac{1}{(2 \pi)^2}
\int_0^\infty
 \lambda d \lambda
 \int_0^{2\pi} d \phi e^{-i \lambda \vec x \vec \xi}
\int_{-\infty}^{\infty} dp e^{i \lambda p} R(p,\vec \xi)
\ee

This formula already expresses the function $f(x,y)$
in terms of its Radon transform 
and the rest of derivation 
consists of its simplification. 
The latter starts with the change of variables $\phi' = \phi,
p' = p - \vec x \vec \xi$ 
 
\be{Fourcont}
f(\vec x)=
\frac{1}{(2 \pi)^2}
\int_0^\infty \lambda d \lambda 
\int_{-\infty}^{\infty}
dp'  e^{i \lambda p'}
 \int_0^{2\pi} d \phi' R(p'+\vec \xi \vec x, \vec \xi)
\equiv \\ \nonumber
\frac{1}{2 \pi}
\int_0^\infty \lambda d \lambda 
\int_{-\infty}^{\infty}
dp  e^{i \lambda p}
\bar R(p,\vec x).
\ee
Here $\bar R(p,\vec x)$ is, obviously, the average of the Radon transform 
over all the straight lines, tangent to the circle of radius 
$p$ and center in the actual point $\vec x$. 

 \be{aver}
\bar R(p,\vec x)=
\frac{1}{2 \pi}
\int_0^{2\pi} d \phi R(p+\vec \xi \vec x, \vec \xi).                          
\ee
To present the inversion 
formula in its final form, one should use the integration by parts
and substitute the formula for (one-dimensional)
Fourier transform of the function $sign(\lambda)$     

\be{Fourfin}
f(\vec x)=
\frac{1}{4 \pi}
\int_{-\infty}^\infty sign (\lambda) \lambda d \lambda dp  e^{i \lambda p}
\bar R(p,\vec x)=
 \frac{i}{4 \pi }
\int_{-\infty}^\infty sign (\lambda) d \lambda dp  e^{i \lambda p}
\bar R'_p (p,\vec x)= \nonumber \\
-\frac{1}{2 \pi}
\int_{-\infty}^\infty \frac{dp}{p}
\bar R'_p (p,\vec x)=
-\frac{1}{ \pi}
\int_{0}^\infty \frac{dp}{p^2} 
(\bar R(p,\vec x) - \bar R(0,\vec x)),
\ee
where the last equality explores once more the integration by parts and 
the fact, that $\bar R(p,\vec x)= \bar R(-p,\vec x)$.
Recalling the definitions (\ref{ParmH},\ref{aver}), one may express the  
double distributions directly in terms of skewed ones (c.f. \cite{Bel}): 

\be{Hfin}
f(x,y)=-\frac{1}{2 \pi^2}
\int_{0}^\infty \frac{dp}{p^2} \int_0^{2\pi} d \phi |cos \phi|
(H (p/cos \phi+x+y tg \phi,tg \phi) 
- H (x+y tg \phi, tg \phi))=\nonumber \\
=-\frac{1}{2 \pi^2}
\int_{-\infty}^\infty \frac{dz}{z^2} 
\int_{-\infty}^\infty
{d\xi}
(H (z+x+y \xi,\xi) 
- H (x+y \xi,\xi)).
\ee
As one clearly see, to recover the double distribution 
one indeed should know the skewed distribution in an 'extended' region 
$-\infty < z, \xi < \infty$. The values $\xi<0$ and 
$,|z|,|\xi|>1$ are described  by the crossing and symmetry relations. 
(\ref{contH},\ref{symH}). it is easy to check by numerical integrations, 
that GPD obtained from various model DD \cite{Rad97} may be recovered with
the reasonable accuracy.  


As it often happened before in the history of the various 
Radon transform applications 
\cite{Radon3}, some of its ingredients were rediscovered in the 
framework of GPD also. In fact, the basic inversion formula in the form 
of triple integral (similar to (\ref{FourR})) was introduced in the 
framework of GPD by 
A.V.~Radyushkin \cite{Rad98} (see also \cite{Bel}).   
Also, the polynomiality condition \cite{Rad97} 
is well known in the general framework of Radon transform as 
the Cavalieri conditions \cite{Radon2}.
The term is related to the interesting geometric interpretation, 
so far not mentioned in the context of DD. Consider the 
first ($n=0$) moment. It is nothing else than the volume 
of the figure, limited by the surface $z=f(\vec x)$ and 
the plane $x,y$.
At the same time, $R(\vec \xi, p)$ may be interpreted as 
a surface of the section of this figure by the plane, containing the 
line $p= \vec \xi \vec x$ and $z$ axis. The first moment is then 
expressing the volume of the figure in terms of the surfaces of 
its intersection with the set of parallel planes. Obviously,
the direction of these planes is unimportant, so that the 
moment does not depend on  $\vec \xi$. The contact with the classical 
Cavalieri principle may be achieved by the particular choice 
of $f(\vec x)=1$ everywhere in the region where it is defined. 
In that case one is dealing with the surface of this region,
represented by the lengths of the set of parallel lines. 
Again it does not depend on their direction and $\vec \xi$. 

\section*{\normalsize \bf Polyakov-Weiss terms and two-dimensional
magneto-statics}

Let us note that separation of the functions $F$ and $G$ is, generally 
speaking, impossible, as soon as  only the information about leading 
twist SPD (GDA) is available.  
Indeed, using the integration by parts and assuming that DD 
turn to zero at the boundary,  
one can absorb the factors $z \cdot P, z \cdot \Delta $ 
to the definitions of DD and 
rewrite (\ref{ParmH}) 
in the following way:    
\be{Parmm}
\langle p'| \bar \psi\biggl(-\frac{z}{2}
\biggr) z \cdot \gamma 
\psi\biggl(\frac{z}{2}\biggr)|p\rangle&=&
-i \int_{-1}^1 dx 
\int_{|x|-1}^{1-|x|} dy e^{-ix Pz - i y \Delta z/2 } N (x,y) 
\\ 
\label{eff}
N (x,y)=\frac{\partial F(x,y)}{\partial x}+
 \frac{\partial G(x,y)}{\partial y},
\ee 
One may absorb the factor  $z \cdot P$ to the definition of the 
SPD in the analogous way. As a result, 
the 'effective' DD $N(x,y)$ is related by the Radon transform to the 
'effective' SPD, which is just the derivative
$\frac{\partial H(z,\xi)}{\partial z}$
of standard SPD. 
Consequently, it is only possible to recover the effective DD, 
while separation of the individual contributions from $F$ and 
$G$ is, generally speaking, impossible.  

It is interesting, that this ambiguity is completely analogous 
to the one arising in the problem of the recovery of the 
electromagnetic vector potential from the magnetic field 
(with a zero total flux).
Indeed, denoting $G(x,y)=A_x, F(x,y)=-A_y, N(x,y)=B_z$,
we can write (\ref{eff}) as $\vec B = rot \vec A$.
Consequently, the effective distribution $N(x,y)$ does not change 
when $F$ and $G$ undergo the following 'gauge transformation'

\be{gauge}
F(x,y) \to F(x,y)+\frac{\partial\alpha(x,y)}{\partial y}, \nonumber \\
G(x,y) \to G(x,y)-\frac{\partial\alpha(x,y)}{\partial x}, \nonumber \\
\alpha(x,y)=-\alpha(x,-y),
\ee 
where the last equation follows from T- (or C-) invariance [\ref{symD}).
One may try to choose 
\be{gauge1}
\alpha(x,y)=\int^x dt G(t,y), 
\ee 
so that the $G$ term is completely eliminated. 
In order to preserve the symmetry properties (\ref{symI}) 
after the gauge transformation, 
the boundary conditions in (\ref{gauge1}) should be chosen in a following way:
\be{gaugsym}
\alpha (x,y)=\frac{1}{2}\biggl(\int_{|y|-1}^x dt G(t,y)
-\int_x^{1-|y|}dt G(t,y)\biggr). 
\ee 
However, these boundary 
conditions are the zero ones in the points $x=\pm(1-|y|)$, 
which in this case gurantees the preservation of zero boundary conditions
for $F$ and $G$,
only if 
\be{bound}
\int_{|y|-1}^{1-|y|} dx G(x,y) =0, 
\ee 
which is indeed true for the case of $I=1$ due to the symmetry properties
(\ref{symI}).
Consequently. in this case, when function $G$ is not required by the 
polynomiality \cite{PW},
it may be completely eliminated by the gauge transformation.
At the same time, for $I=0$ it is, generally speaking,
impossible to assume simultaneously
the zero boundary conditions at 
 $x=1-|y|$ and   $x=|y|-1$ and eliminate $G$ completely.
However, it is possible to reduce $G(x,y)$ to the function 
of {\it one} variable: 
\be{bound1}
G(x,y) \to d(y)=\frac{1}{2(1-|y|)}\int_{|y|-1}^{1-|y|} dx G(x,y), 
\ee 
by the following choice of the gauge:
\be{gaugepw0}
\alpha (x,y)=\frac{1}{2}\biggl(\int_{|y|-1}^x dt G(t,y)
-\int_x^{1-|y|}dt G(t,y)\biggr)- x d(y). 
\ee 
Here the $x$-independent contribution to $G(x,y)$ is  provided by 
the last term which guarantees the zero boundary condition,
One can change in this boundary-correcting term 
the linear function to the step one, 
\be{gaugepw}
\alpha_{PW}(x,y)=\frac{1}{2}\biggl(\int_{|y|-1}^x dt G(t,y)
-\int_x^{1-|y|}dt G(t,y) - sign(x) D (y)\biggr) \nonumber \\=
\theta (x<0)\int_{|y|-1}^x dt G(t,y)-\theta (x>0)\int_x^{1-|y|}dt G(t,y), 
\ee
where $D(y)$ determines  
the resulting expression for $G$ which now resides at $x=0$:
\be{bound2}
G(x,y) \to \delta (x) D(y)
=\delta(x)\int_{|y|-1}^{1-|y|} dt G(t,y). 
\ee 
The existence of such a gauge 
transformation justifies the original suggestion \cite{PW}
to consider this special form of $G(x,y)$. 

\section*{\normalsize \bf Discussion and Conclusions}

We studied crossing properties of Generalized Parton Distributions
and showed, that Ra\-dyush\-kin's Double Distributions 
express them in the most simple way. Namely, they simultaneously 
describe, up to $t$ and $W^2$ dependence, the DVCS and 
two-photon production of hadron pairs. This allows, in principle,
to recover them from the data in the both channels, making use 
of the inverse Radon transform, solving therefore a problem of Radon 
tomography in the field of non-perturbative QCD. 
We also considered the ambiguity due to the Polyakov-Weiss terms
and showed, that it is completely analogous for the gauge ambiguity 
of the vector potential of the static two-dimensional magnetic field. 
The observable quantity is in this sense 'effective' DD $N(x,y)$,
which may be decomposed to the the $F$ and $G$ structures, adopting the 
particular gauge condition.  

The development of the method may, first of all, include the more 
refined versions of Radon transform (see e.g. numerous Refs. in 
\cite{Radon3}),  
which would allow to use the limited range of angles of Radon
projections (limited range of $\xi$). In that case, one may 
essentially limit the required input, and in particular,
avoid the use of both channels. Also, one might 
include the possibilities of the singularities of 
recovered functions \cite{Radon4}.
Other natural development would be to include the twist 3 
contributions \cite{tw3}. 
Another interseting opportunity may be provided
by the process  $\gamma^\ast \gamma \to 3 \pi$, the GDA \cite{3pi}
for which depend on the {\it two} light-cone fractions. 
The correspondent 'Triple Distributions' would 
depend on three variables, so to recover 
them one should consider the tomography in three-deimensional space, 
which (like for any odd-dimensional space)
is known \cite{Radon2,Radon3} to have the local form due to the 
Hyugens principle. 

Moreover, the Radon transform might be useful in the more 
general context of quantum field theory. Especially interesting 
seems to be its application to the general problem 
of crossing invariance, as it allows to consider (instead of functions 
in the {\it different} regions of their variables) the different 
projections of the {\it common} function, as it was discussed here 
in the particular case of GPD. 

Another interesting application may arise from the fact, that 
inverse Radon transform naturally recovers information about the 
interior of some region, which makes a contact to the 
famous holographic principle \cite{hol}. One could add here,
that Radon transform technique is known for a years to be one of the 
main tools in the field of {\it optical} holography \cite{Radon3}.     

\section*{\normalsize \bf Acknowlegments.}

I would like to thank  A.V.~Radyushkin and C.~Weiss
for the enlightening discussions of the various aspects  
of the double distributions.
I am also thankful to M.~Diehl and D.~M\"uller 
for useful discussions and comments. 
This work was partially supported by DFG, RFFI grant 00-02-16696
and INTAS Project 587 (Call 2000).


\begin{thebibliography}{99}
\bibitem{DVCS1}
D. M\"uller, D. Robaschik, B. Geyer,
F.M. Dittes,J. Horejsi, Fortschr. Phys. {\bf 42} (1994) 101.
\bibitem{DVCS2}
X. Ji, Phys. Rev.{\bf D55} (1997) 7114.
\bibitem{exp1}
P.~R.~Saull  [ZEUS Collaboration],
``Prompt photon production and
observation of deeply virtual Compton  scattering,''
hep-ex/0003030.
\bibitem{exp2} Rainer Stamen [H1-Collaboration],
`` Measurement of the Deeply Virtual Compton Scattering at Hera'',
H1prelim-00-17, DIS 2000 and IHEP 2000. 

\bibitem{DGPT} M. Diehl, T. Gousset, B. Pire and O.V. Teryaev,
Phys.\ Rev.\ Lett.\ {\bf 81} (1998) 1782.

\bibitem{DGP} M. Diehl, T. Gousset, B. Pire,
 Phys Rev {\bf D62} (2000) 073014.

\bibitem{fac} A. Freund, Phys.Rev. {\bf D61} (2000) 074010.

\bibitem{BL} G.P. Lepage and S.J. Brodsky, Phys.\ Rev.\ {\bf D22}
(1980) 2157.

\bibitem{RR} S. Ong, Phys.\ Rev.\ {\bf D52}, (1995) 3111;
R. Jakob, P. Kroll and M. Raulfs, J. Phys. {\bf G22}, (1996) 45;
P. Kroll and M. Raulfs, Phys.\ Lett.\ {\bf B387}, (1996) 848;
A.V. Radyushkin and R.T. Ruskov, Nucl.\ Phys.\ {\bf B481}, (1996) 625;
I.V. Musatov and A.V. Radyushkin, Phys.\ Rev.\ {\bf D56}, (1997) 2713.

\bibitem{PW}
M.V. Polyakov and C. Weiss, Phys. Rev. {\bf D60} (1999) 114017.

\bibitem{Rad97}
A.~V.~Radyushkin,
Phys.\ Rev.\  {\bf D56} (1997) 5524.

\bibitem{ET84}
A.V. Efremov, O.V. Teryaev, Sov. Journ. Nucl. Phys. {\bf 39} (1984) 962.

\bibitem{Radt}
A.~V.~Radyushkin,
Phys.\ Rev.\  {\bf D58} (1998) 114008.

\bibitem{Radon1} J.~Radon, Berichte S\"achsische Akademie 
der Wissenschaften, Leipzig, Math-Phys. Kl., {\bf 69} (1917) 262.
 
\bibitem{Radon2} I.M.~Gel'fand, M.I.~Graev and N.Ya.~Vilenkin, 
{\it Generalized functions}, V.5, Academic Press, N.Y.-London, 1966;
I.M.~Gel'fand, S.G.~Gindikin and M.I.~Graev, {\it Selected problems of 
integral geometry (in Russian)}, Moscow, Dobrosvet, 2000. 

\bibitem{Radon3} S.~Deans, {\it The Radon transform and some 
of its applications}, Wiley-Interscience, 1983. 

\bibitem{Radon4} D.A.~Popov, Russian Math.Surveys, {\bf 53} (1998) 109.

\bibitem{Bel} A.~V.~Belitsky, D.~Muller, A. Kirschner and A. Sch\"afer,
hep-ph/0011314. 

\bibitem{Rad98}
A.~V.~Radyushkin,
Phys.\ Lett.\  {\bf B449} (1999) 81.

\bibitem{tw3}
I.~V.~Anikin, B.~Pire and O.~V.~Teryaev,
Phys Rev {\bf D62} (2000) 071501;
M.~Penttinen, M.~V.~Polyakov, A.~G.~Shuvaev and M.~Strikman,
 Phys. Lett. {\bf B491} (2000) 96;
A.~V.~Belitsky and D.~Muller,
Nucl.Phys. {\bf B589} (2000) 611; 
J.~Bl\"umlein, B.~Geyer, M.~Lazar and
D.~Robaschik, Nucl.\ Phys.\ Proc.\ Suppl.\  {\bf 89} (2000) 155;
A.~V.~Radyushkin,~C.~Weiss, Phys.Lett. {\bf B493} (2000) 332;  hep-ph/0010296.

\bibitem{3pi}  B.~Pire and O.~V.~Teryaev,
Phys.Lett. {\bf B496} (2000) 76. 

\bibitem{hol} G.~t'Hooft, gr-qc/9321026.  

\end{thebibliography}
\end{document}